\documentclass[preprint,preprintnumbers,amsmath,amssymb,groupedaddress,endfloats]{revtex4}

\usepackage{graphicx}
\usepackage{dcolumn}
\usepackage{bm}

\begin{document}

\title{Resonant tunneling diode with spin polarized injector}

\author{A. Slobodskyy}
\affiliation{%
Physikalisches Institut (EP3), Universit\"{a}t W\"{u}rzburg, Am
Hubland, D-97074 W\"{u}rzburg, Germany
}%
\author{C. Gould}%
\affiliation{%
Physikalisches Institut (EP3), Universit\"{a}t W\"{u}rzburg, Am
Hubland, D-97074 W\"{u}rzburg, Germany
}%
\author{T. Slobodskyy}%
\affiliation{%
Physikalisches Institut (EP3), Universit\"{a}t W\"{u}rzburg, Am
Hubland, D-97074 W\"{u}rzburg, Germany
}%
\author{G. Schmidt}%
\affiliation{%
Physikalisches Institut (EP3), Universit\"{a}t W\"{u}rzburg, Am
Hubland, D-97074 W\"{u}rzburg, Germany
}%
\author{L.W. Molenkamp}%
\affiliation{%
Physikalisches Institut (EP3), Universit\"{a}t W\"{u}rzburg, Am
Hubland, D-97074 W\"{u}rzburg, Germany
}%

\author{D. S\'{a}nchez}%
\affiliation{Departament de F\'{\i}sica, Universitat de les Illes
Balears, E-07122 Palma de Mallorca, Spain}
\affiliation{D\'epartement de Physique Th\'eorique, Universit\'e de
Gen\`eve, CH-1211 Gen\`eve 4, Switzerland}

\date{\today}

\begin{abstract}
We investigate the current--voltage characteristics of a II-VI
semiconductor resonant-tunneling diode coupled to a diluted
magnetic semiconductor injector. As a result of an external
magnetic field, a giant Zeeman splitting develops in the
injector, which modifies the band structure of the device,
strongly affecting the transport properties.
We find a large increase in peak amplitude
accompanied by a shift of the resonance to higher voltages with
increasing fields. We discuss a model which shows that the effect
arises from a combination of three-dimensional incident
distribution, giant Zeeman spin splitting and broad resonance
linewidth.
\end{abstract}

\maketitle

One of the interesting designs for eventual spintronics devices are
structures based on magnetic semiconductor resonant tunneling diodes
(RTDs) which have promising futures as spin filters and spin
aligners. However, before these structures can make their way into
actual devices, we must first understand the basic behaviors of
these magnetic structures and how they differ from their
non-magnetic counterparts. Early studies into magnetic semiconductor
tunneling structures have reported spin injection from either
paramagnetic or ferromagnetic semiconductor \cite{fidi,aw} into
non-magnetic light emitting diodes. A primary tradeoff between these
two options is that while ferromagnets allow for non-volatile
operation without an external magnetic field, they can thus far only
be made p-type, and the short mean free path of holes compared to
electrons significantly reduces coherent transport phenomena. Using
paramagnetic semiconductors, we have previously explored the
behavior of RTDs with a magnetic quantum well by optical \cite{waag}
and electrical means, \cite{slo03} and now turn our attention to
RTDs fitted with magnetic injectors. Here, the polarization of spin
originates in the bulk behavior of the injector, and the RTD itself
is essentially a non-magnetic device which is fed with a polarized
source of electrons. As a magnetic field increases, we observe
important changes in the transport characteristics of the RTD
stemming from the realignment of the band profile in the device, and
explain how this can be understood in the framework of a
Landauer-B\"uttiker--type model. A correct understanding of this
mechanism is therefore essential before proceeding to more
sophisticated device structures combining spin injection, detection
and manipulation.

\begin{figure}
\centerline{\includegraphics[width=3.375in]{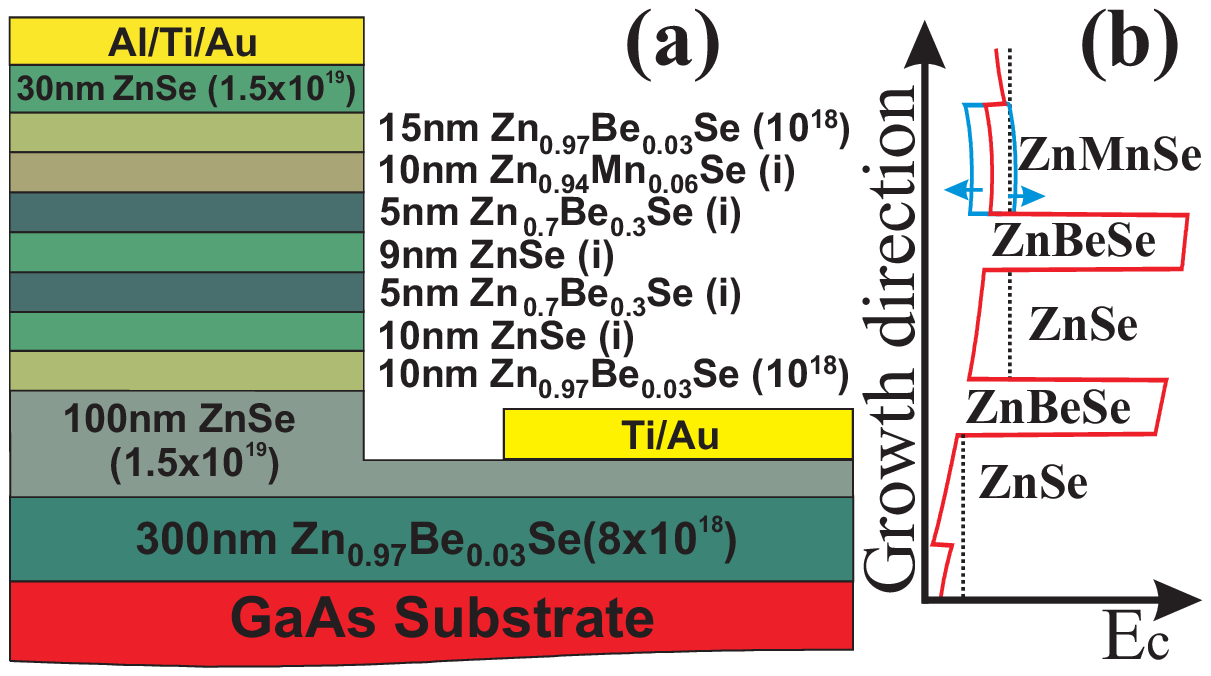}}
\caption{\label{mbe} a) Layer structure of the device and b)
schematic view of resonance tunnel diode band structure under an
applied bias.}
\end{figure}

The studied sample is an all-II-VI semiconductor resonant tunneling diode
structure with a conduction band profile similar to that used
in previous work on II-VI and III-V\cite{slo03} based RTD and fitted
with a dilute magnetic semiconductor (DMS) injector layer. The
tunneling region consist of a 9 nm thick ZnSe quantum well (QW)
sandwiched between two 5 nm thick Zn$_{.7}$Be$_{.3}$Se barriers.
Below the tunneling region is a 10 nm thick ZnSe collector layer.
The injector is a 10 nm thick Zn$_{.94}$Mn$_{.06}$Se layer used
to inject spin polarized electrons into the diode. The remaining
layers are needed to ensure a proper doping profile and to allow for
the fitting of high quality ohmic contacts. Full details of the
layer stack are given in Fig.~\ref{mbe}(a).

The contact resistance of the devices is kept to a minimum by using
an in-situ Al(10 nm)/Ti (10 nm)/Au (30 nm) top contact, while the
ex-situ bottom contact is fabricated by etching down to the very
highly doped ZnSe layer, and using large area (500$^{2} \mu
$m$^{2})$ Ti-Au contact pad. The sample is patterned in 100
$\mu$m$^{2}$ pillars using standard optical lithography, wet each
and lift-off.

Measurements are performed in $^4$He bath cryostat equipped with a high field
superconducting magnet using standard DC transport techniques. Care
was taken to construct a circuit with a low (40 $\Omega$) resistor in
parallel to the diode to prevent the diode from going into
oscillations in the negative differential conductance region.
\cite{Leadb}. Current measurements consist of measuring the voltage drop
across a relatively small 30 Ohm series resistor.

A schematic of the band profile is shown in Fig.~\ref{mbe}. Both the injector and collector sides of the
tunneling structure are gradient doped in order to insure a relatively low Fermi energy at the point of injection, and thus a sharp resonance \cite{Wie}. Under the influence of an external magnetic
field, the DMS injector exhibits a giant Zeeman splitting of up to 20 mV, and the
degeneracy of its spin states is lifted following a Brillouin
function \cite{Gaj}, creating a spin polarized carrier population
via the transfer of electrons from the higher energy spin band to
the lower one. In order to maintain proper alignment of the Fermi
energy, this imbalance in the population of the two spin species
must lead to a splitting of the bottom of the spin up and spin down
conduction bands, leading to different band profiles for each of the
spin species in the injector region, and indicated in the diagram.

Current-voltage characteristics for the device for different
amplitudes of perpendicular-to-plane magnetic fields ($B$) are shown
in Fig. 2. At zero field, the sample exhibits typical RTD behavior,
showing a strong resonance peak at 36 mV with a peak to valley ratio
just over 1. The additional resonance visible at ~96 mV in the $B=0$
curve is the well known LO phonon replica \cite{Leadb} which is
separated from the direct resonance by the energy of the LO phonon
of the well material, and can be used to calibrate the voltage scale
to the energy of the levels in the QW. A resonance associated with
the second well level occurs at around 0.3 V (not shown in the
figure). Because of the intrinsic asymmetry of the device due to the
presence of the GaAs substrate, the resonances for the negative bias
direction are not as pronounced. We confirmed the absence of
charging effects in the device by verifying that I-V curves for
different sweep direction were identical \cite{Martin}.

As a magnetic field is applied, the peak shifts to higher voltages
and strongly \textit{increases} in amplitude. The strongest
dependence of the peak amplitude and position on the magnetic field
is observed at low fields, with the behavior saturating before 10 T.
The behavior is therefore consistent with the Brillouin like
behavior expected from the giant Zeeman splitting of the injector.
Note also that the phonon replica causes a higher energy echo of the
QW state, but since the effects we are studying result from
properties of the injector, the phonon replica peak simply follows
the magnetic field dependance of the main resonance.

Figure 3 shows the temperature dependence of the 2~T I-V curve for
temperatures from 1.3 to 8~K. The $B=0$ curve, which is temperature
independent in this range, is included for comparison. It is clear
that the increase in temperature has an identical effect on the I-V
curves as would a lowering in the magnetic field. Indeed, we
verified that the only effect of temperature is to rescale the
effect of the magnetic field on the Zeeman splitting in the injector
exactly as would be expected from the fact that the argument of the
Brillouin function scales as $B/(T+T_{\rm eff})$ as described below.

The slight shift of the peak to higher bias with increasing magnetic
field is fairly intuitive and can be directly expected from the
schematic of Fig. 1b. It is well established that the maximum of the
resonance occurs when the bottom of the conduction band is brought
into alignment with the well level \cite{lur85}. As the field
increases, the peak becomes more and more dominated by the majority
spin conduction band. Since the bottom of this band is moving to
lower energies, a higher bias will be required to bring this band in
alinement with the well level, producing a shift of the resonance
towards higher bias.

The strong increase in the current amplitude is similarly related to
the change of the conduction band energy with magnetic field $B$,
but its manifestation is somewhat more subtle. Let $E_\sigma=E_c+s
h/2$ denote the spin-split band bottom with $h$ the spin splitting
[$s=+$($-$) for $\sigma=\uparrow$($\downarrow$)]. Due to the
exchange interaction between localized Mn ions and band electrons a
giant Zeeman splitting arises in the DMS which is known to be given
by:
\begin{equation}\label{eq_h}
h=N_0 \alpha x S_0 B_S[S g \mu_B B/k_B(T+T_{\rm eff})]\,,
\end{equation}
where $N_0\alpha$ is the exchange integral, $x$ the Mn concentration
of $S=5/2$ Mn spins, $g$ is the Lande factor, $B_s$ the Brillouin
function, and $S_0$ and $T_{\rm eff}$ are the Mn effective spin and
temperature, respectively. One might naively expect that a spin
splitting of the conduction band leaves the DMS injector electron
density constant (keeping $E_F$ fixed). This is indeed correct for a
2D injector~\cite{san02} since the spin-dependent density
$n_\sigma=(m/2\pi\hbar^2) (E_F-E_\sigma)$ is independent of energy
because the density of states is constant. However, for a 3D system
$n_\sigma$ depends {\em nonlinearly} on $h$, so that (as we will
show below) the total electron density $n=n_\uparrow+n_\downarrow$
is an {\em increasing} function of $h$ provided $E_F$ is kept fixed.
Since the current peak is mainly determined by $n$, it follows from
Eq.~(\ref{eq_h}) that the peak amplitude increases with $B$.
Therefore, the current increase effect has a {\em geometrical}
origin.

To gain further insight, we perform numerical simulations of
the electron transport along the $z$-direction
[growth direction, see Fig.~\ref{mbe}(b)].
Because a similar peak amplitude
increase is also observed when $B$ is an in-plane
magnetic field indicating that the effect is independent of the direction of the magnetic field, orbital effects are not taken into account.
In addition, we neglect spin relaxation effects
and the current density
$J=J_\uparrow+J_\downarrow$ is thus carried by both spin species in parallel,
\begin{eqnarray}\label{eq_cur}
J_\sigma&=&\frac{em}{4\pi^2\hbar^3}
\int_{eV+s h/2}^\infty dE_z dE_\bot \, T_\sigma(E_z,\varepsilon_0,V) \\ \nonumber
&\times& [f_L(E_z+E_\bot)-f_R(E_z+E_\bot)]\,,
\end{eqnarray}
with $V$ the bias voltage applied to the structure. While this
assumptions limits the model to the case of coherent transport, the
good agreement it shows with experience suggests that this assumption
is reasonable for the device in question. We take the spin splitting
into account in the longitudinal energy $E_z$. The Fermi functions
$f_L$ and $f_R$ describe the distribution of electrons with total
energy $E_z+E_\bot$ in the left and right leads with electrochemical
potentials $\mu_L=E_F$ and $\mu_L=E_F-eV$, respectively. Band-edge
effects are incorporated in the lower limit of the integral.

\begin{figure}
\centerline{\includegraphics[width=3.375in]{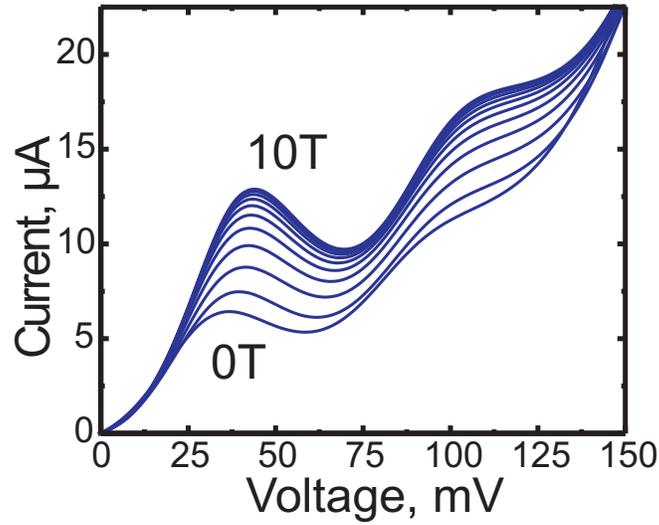}}
\caption{\label{shrift} Magnetic field dependence from 0 to 10T in
1T steps, of the I-V curves of the first resonance at 4.2~K.}
\end{figure}

In Eq.~(\ref{eq_cur}), the transmission $T$ conserves the momentum
parallel to the interfaces and depends, quite generally, on $E_z$,
$V$ and the sub-band bottom energy $\varepsilon_0$ of the quantum
well. Close to resonance, it is a good approach to take a Lorentzian
shape,
$T_\sigma=\Gamma_\sigma^L\Gamma_\sigma^R/[(E_z-\varepsilon_0)^2+\Gamma^2/4]$,
where $\Gamma_\sigma^L$ ($\Gamma_\sigma^R)$ is the partial decay
width due to coupling to lead L (R) and $\Gamma=\Gamma_L+\Gamma_R$
the total broadening per spin. We consider symmetric barriers for
simplicity but note that it is important in the strongly nonlinear
regime (i.e., around the current peak) to take into account the
energy (and voltage) dependence of the tunneling rates\cite{bla99}.
The barrier height is given by the conduction band offset between
ZnSe and Zn$_{0.7}$Be$_{0.3}S$e, which is approximately $0.6$~eV.
Using ZnSe parameters we find for the first resonant level
$\varepsilon_0=21$~meV. Spin saturation takes place at around
$h=20$~meV, thereby $E_F=10$~meV. This implies that for small
resonance width the onset of the $I$--$V$ curves of
Fig.~\ref{shrift} should occur at $\varepsilon_0-E_F=11$~meV. This
fact, together with the low peak-to-valley ratio that the $I$--$V$
curves exhibit, suggests a large value of $\Gamma$, probably due to
disorder or interface roughness. Therefore, we set $\Gamma=15$~meV
at $V=0$ and $h=0$ and fit the peak current at $B=0$ with the
experimental value.

We show the results in Fig.~\ref{theory} from $h=0$ to $h=2E_F$. The
$I$--$V$ curves reproduce the experimental observations and the
agreement is fairly good. The negative differential conductance
region extends from the resonance peak because we consider only the
first resonant level and neglect non-resonant contributions to the
tunneling current. In Fig.~\ref{theory} the voltage axis is scaled
with a lever arm $\mathcal{L}=2$, consistent with that extracted
from the position of the phonon replica. Since the broadening is a
large energy scale, we can expand Eq.~(\ref{eq_cur}) at zero
temperature in powers of $1/\Gamma$. We find for small spin
splitting that the resonance peaks at $eV_{\rm
res}/\mathcal{L}=\varepsilon_0-E_F/3+h^2/3 E_F$ clearly shows the
shift of $V_{\rm res}$ with increasing $h$. Substituting the
parameters, we obtain $V_{\rm res}=37$~mV at $h=0$, in accordance
with Fig.~\ref{shrift}. Moreover, inserting this result in
Eq.~(\ref{eq_cur}) we find to leading order in $1/\Gamma$ that the
peak current $J_p=(em/2\pi^2\hbar^3) E_F^2 (1+h^2/4E_F^2)$ is indeed
an increasing function of the magnetic field. In the inset of
Fig.~\ref{theory} we present the current peak as a function of $h$,
showing a remarkable agreement with the experimental observations.
We note that since the only experimental effect of varying the
temperature is a rescaling of $h$, this model is equally valid for
the temperature dependent data. For comparison, we provide the
current increase when the energy dependence of $\Gamma$ is
neglected, which qualitatively agrees with our results, indicating
that the simple model~(\ref{eq_cur}) captures the key points of the
experiment. Conceptually, this result can be viewed as resulting
from the change in the position of the conduction band in the
injector which is required to maintain the alignment of the Fermi
level throughout the device while allowing for the magnetic field
induced redistribution of spin populations. This pushes down the
bottom of the majority spin conduction band, which thus requires a
higher bias to be brought into alignment with the well level, and
consequently moves the maximum of the resonance to higher bias.
Additionally, the redistribution of the spin population increases
the size of the Fermi sphere of the majority band, providing more
states at the Fermi energy available for tunneling, and thus
increasing the current.

Indeed, the results we obtain are reminiscent of Ref.~\onlinecite{cho92}, where
the In content of a GaInAs emitter in a III-V RTD was varied.
As a consequence, the band alignment changes similarly to our present structure, and the peak current
increases. In our case, the increase is
due exclusively to the giant spin splitting in the injector.
Thus, the current increase can be tuned with a magnetic field
without changing the sample parameters.

In conclusion, we have presented $I$--$V$ characteristics
of a II-VI resonant-tunneling diode attached to a diluted magnetic
semiconductor injector. We have observed both an enhancement
and a voltage shift of the resonance current peak when the applied
magnetic field increases.
The results are consistent with a giant
Zeeman splitting in the injector since the current saturates
at a few Tesla and the temperature dependence follows
the magnetization of a paramagnetic system. The Zeeman splitting induced redistribution of spin carriers in the injector leads to a modification of the conduction band structure, causing a lowering of the majority spin conduction band and an increase of available states for tunneling, which in turn is responsible for the changes in transport properties. We have discussed a transport model which replicates most
of the features seen in the experiment. Our findings
offer the unique possibility of producing high peak currents
which arise from spin effects only.

{\em Acknowledgments.---}The authors thank M.~B\"uttiker for useful
discussions and V. Hock for sample fabrication, as well as Darpa
SpinS, ONR, SFB 410, BMBF, RTN, and the ``Ram\'on y Cajal'' program
for financial support.

\pagebreak

\pagebreak

\begin{figure}
\centerline{\includegraphics[width=3.375in]{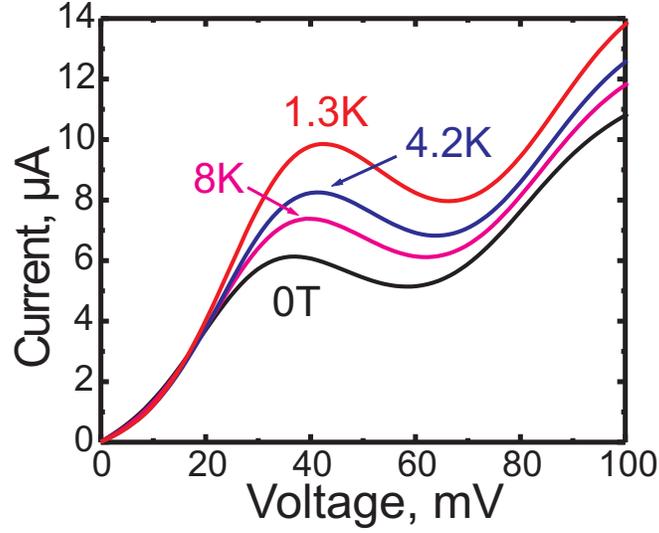}}
\caption{\label{temperature} Temperature dependence of the
resonance.
}
\end{figure}

\begin{figure}
\centerline{\includegraphics[width=3.375in]{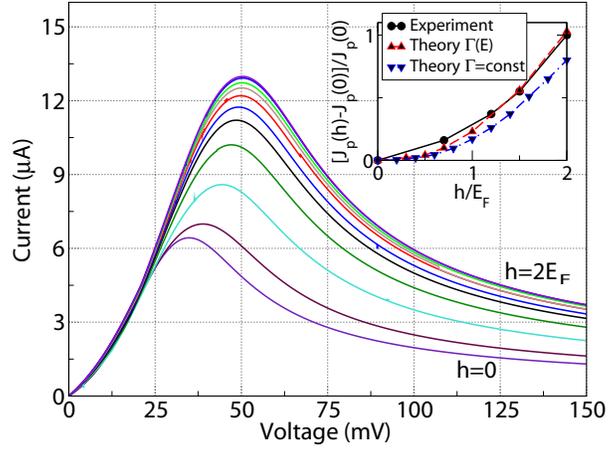}}
\caption{\label{theory} Theoretical $I$--$V$ curves at 4~K for a RTD
with a spin-polarized injector increasing the spin splitting from
$h=0$ to $h=2 E_F$ in steps of $h=0.2 E_F$ (from bottom to top).
Inset: Normalized current peak $J_p$ as a function of $h$.
Experimental values taken from the data of fig. 2 and model data are
shown for comparison.}
\end{figure}


\end{document}